\newcommand{\AP}[3]{Ann.\ Phys.\ {\bf #1},\ #2 (#3)}
\newcommand{\NPA}[3]{Nucl.\ Phys.\ {\bf A#1},\ #2 (#3)}
\newcommand{\NPB}[3]{Nucl.\ Phys.\ {\bf B#1},\ #2 (#3)}

\newcommand{\PLB}[3]{Phys.\ Lett.\ B\ {\bf #1},\ #2 (#3)}
\newcommand{\PR}[3]{Phys.\ Rep.\ {\bf #1},\ #2 (#3)}
\newcommand{\PRL}[3]{Phys.\ Rev.\ Lett.\ {\bf #1},\ #2 (#3)}
\newcommand{\PRA}[3]{Phys.\ Rev.\ A\ {\bf #1},\ #2 (#3)}

\newcommand{\PRC}[3]{Phys.\ Rev.\ C\ {\bf #1},\ #2 (#3)}
\newcommand{\PRD}[3]{Phys.\ Rev.\ D\ {\bf #1},\ #2 (#3)}
\newcommand{\JPG}[3]{J.\ Phys.\ G\ {\bf #1},\ #2 (#3)}

\newcommand{\ZPC}[3]{Z.\ Phys.\ C\ {\bf #1},\ #2 (#3)}

\newcommand{\EPJA}[3]{Eur.\ Phys.\ J.\ A\ {\bf #1},\ #2 (#3)}
\newcommand{\PTP}[3]{Prog.\ Theo.\ Phys.\ {\bf #1},\ #2 (#3)}

\renewcommand\k{\kappa}

\newcommand{\diracslash}[1]{#1\llap{/\kern2pt}}

\newcommand{\be}{\begin{equation}}
\newcommand{\ee}{\end{equation}}
\newcommand{\bea}{\begin{eqnarray}}
\newcommand{\eea}{\end{eqnarray}}
\newcommand{\ba}[1]{\begin{array}{#1}}
\newcommand{\ea}{\end{array}}

\documentclass[prd,aps,floats,nofootinbib,showpacs,floatfix]{revtex4}
\usepackage{graphicx}
\begin{document}
%13.8.08

\title{Kaon and antikaon optical potentials in isospin asymmetric 
hyperonic matter}

\author{Amruta Mishra}
\email{amruta@physics.iitd.ac.in}
\affiliation{Department of Physics,Indian Institute of Technology,Delhi,
Hauz Khas, New Delhi - 110 016, India}

\author{Arvind Kumar}
\email{iitd.arvind@gmail.com}
\affiliation{Department of Physics,Indian Institute of Technology,Delhi,
Hauz Khas, New Delhi - 110 016, India}

\author{Sambuddha Sanyal}
\email{sambuddhasanyal@gmail.com}
\affiliation{Department of Physics,Indian Institute of Technology,Delhi,
Hauz Khas, New Delhi - 110 016, India}

\author{Stefan Schramm}
\email{schramm@th.physik.uni-frankfurt.de}
\affiliation{
Frankfurt Institute for Advanced Studies,
J.W. Goethe Universit\"at,
Ruth-Moufang-Str. 1, D-60438 Frankfurt am Main, Germany}

\begin{abstract}
The medium modifications of the energies of kaons and antikaons in 
isospin asymmetric hyperonic matter are investigated using a chiral SU(3)
model. The isospin dependent medium effects, are important for asymmetric 
heavy ion collision experiments, as well as relevant for the neutron star 
phenomenology as the bulk matter in the interior of the neutron star is 
highly isospin asymmetric. The effects of hyperons on the medium 
modifications of the kaons and antikaons in the strange hadronic matter 
are investigated in the present work and are seen to be appreciable 
for hadronic matter with large strangeness fractions. The study of 
the K-mesons in the asymmetric strange hadronic matter can be especially 
relevant for the compressed strange baryonic matter which can result 
from asymmetric heavy ion collision experiments in the future 
accelerator facility FAIR at GSI.

\end{abstract}

%\date{ }

\maketitle

%-----------------------------------------------------------------------
\section{Introduction}

The topic of study of the in-medium properties of hadrons 
is an important problem in strong interaction physics, which has
relevance in the high energy heavy-ion collision 
experiments, as well as in neutron star phenomenology.
In heavy ion collision experiments, the medium modifications of the hadrons
can be seen in different observables, like particle yield, particle
spectra as well as in their collective flow. The study of medium modifications 
of K-mesons were initiated by Kaplan and Nelson \cite{kaplan},
who suggested the possibility of antikaon condensation in the interior
of the neutron stars due to the drop in the mass of the antikaons
in the nuclear medium.  However, recent experimental observations 
on neutron star phenomenology impose constraints on the nuclear
equation of state (EOS). The EOS for the nuclear matter obtained 
using an efective model should be consistent with the astrophysical 
bounds to be acceptable as an EOS for neutron star matter \cite{blaschke,ozel}.
Recently, the nuclear matter EOS have been investigated consistent with
the neutron star phenomenology as well as data for collective flow
in heavy ion collision experiments \cite{klahn}.
The in-medium  modification of kaon/antikaon
properties can be observed experimentally 
\cite{FOPI,Laue99,kaosnew,Sturm01,Forster02} in 
relativistic nuclear collisions from their abundance, spectra as well as
collective flow. There have also been intense theoretical investigations
\cite{lix,cmko,Li2001,Cass97,brat97,CB99,laura03,Effenber00,Aichelin,Fuchs}
on the kaon and antikaon properties in the dense and hot hadronic 
matter, and to study the effects of such modifications on the 
observables of the high energy heavy ion collision experiments.
Hence there have been extensive research to obtain the in-medium properties 
of kaons and antikaons due to their relevance for the high energy 
heavy ion collision epxeriments. These have been calculated by using 
several methods like extracting from kaonic atom data, using chiral 
lagrangians or by coupled channel methods \cite{kaonopt}. However, 
the kaon/antikaon potential in the hot and dense hadronic matter
remains still an unresolved issue.

The isospin effects in hot and dense hadronic matter \cite{asym}
are important in isospin asymmetric heavy-ion collision
experiments. Within the UrQMD model the density dependence of the
symmetry potential has been studied by investigating observables
like the $\pi^-/\pi^+$ ratio, the n/p ratio, the $\Delta ^-/\Delta ^{++}$ 
ratio, NN scattering cross section and pion flow for 
neutron rich heavy ion collisions \cite{li134}. 
Within this UrQMD model, the isospin effects on the production
of $K^0$ and $K^+$ \cite{li2} have also been studied.

In \cite{haar} it was pointed out that one has to take into account
the effect of the Haar measure in mean field approximations of non-linear
chiral models, which is especially relevant in the high-temperature regime.
An extended discussion
of the model presented here, including this contribution should
be performed in future studies.

In the present investigation we use a chiral SU(3) model
\cite{paper3,kristof1} for the description of hadrons in the medium.
The model has been used to study nuclear matter, finite nuclei,
and hyperonic matter. The properties of the  vector mesons 
have also been investigated in the nuclear medium within this model
\cite{hartree,kristof1}. The energies of kaons
(antikaons) in the (asymmetric) nuclear matter 
\cite{kmeson1,isoamss,isoamss1} were also studied within
this framework and the isospin dependence was seen to be
appreciable for high densities abd hence will be particularly 
relevant in the future compressed baryonic matter (CBM) experiment
at FAIR, GSI. The kaon and antikaon properties have been 
investigated consistent with the low energy KN scattering data
\cite{juergen,cohen}. In the present work, we investigate the effect 
of strangeness on the kaon and antikaon optical potentials in the 
isospin asymmetric hyperonic matter, consistent with the
low energy kaon nucleon scattering lengths for channels I=0 and I=1
\cite{isoamss,isoamss1}.

The outline of the paper is as follows: In section II we shall
briefly review the SU(3) model used in the present investigation.
Section III describes the medium modification of the K($\bar K$)
mesons in this effective chiral model. In section IV, we discuss the
results obtained for the optical potentials of the kaons and
antikaons and the isospin-dependent effects on these optical
potentials in asymmetric hyperonic matter. Section V summarizes 
our results and discusses possible extensions of the calculations.

\section{ The hadronic chiral $SU(3) \times SU(3)$ model }
The effective hadronic chiral Lagrangian used in the present work is given as
\be
{\cal L} = {\cal L}_{kin} + \sum_{ W =X,Y,V,{\cal A},u }{\cal L}_{BW}
          + {\cal L}_{vec} + {\cal L}_0 + {\cal L}_{SB}
\label{genlag} \ee are discussed. Eq. (\ref{genlag}) corresponds
to a relativistic quantum field theoretical model of baryons and
mesons adopting a nonlinear realization of chiral symmetry
\cite{weinberg,coleman,bardeen} and broken scale invariance (for
details see \cite{paper3,hartree,kristof1}) as a description of 
the hadronic matter. The model was used successfully to
describe nuclear matter, finite nuclei, hypernuclei and neutron
stars. The Lagrangian contains the baryon octet, the spin-0 and
spin-1 meson multiplets as the elementary degrees of freedom. In
Eq. (\ref{genlag}), $ {\cal L}_{kin} $ is the kinetic energy term,
$  {\cal L}_{BW}  $ contains the baryon-meson interactions in
which the baryon-spin-0 meson interaction terms generate the
baryon masses. $ {\cal L}_{vec} $ describes the dynamical mass
generation of the vector mesons via couplings to the scalar fields
and contains additionally quartic self-interactions of the vector
fields.  ${\cal L}_0 $ contains the meson-meson interaction terms
inducing the spontaneous breaking of chiral symmetry as well as a
scale invariance breaking logarithmic potential. $ {\cal L}_{SB} $
describes the explicit chiral symmetry breaking.

The baryon-scalar meson interactions generate the baryon masses
and the parameters corresponding to these interactions are adjusted 
so as to obtain the baryon masses as their experimentally measured 
vacuum values. For the baryon-vector meson interaction terms, there
exist the $F$-type (antisymmetric) and $D$-type (symmetric) couplings.
Here we will use the antisymmetric coupling \cite{paper3,isoamss}
because, following the universality principle  \cite{saku69}
and the vector meson dominance model, one can conclude that
the symmetric coupling should be small.
Additionally we choose the parameters \cite{paper3,isoamss} so as to
decouple the strange vector field $
\phi_\mu\sim\bar{s} \gamma_\mu s $ from the nucleon,
corresponding to an ideal mixing between $\omega$ and $\phi$.
A small deviation of the mixing angle from the ideal mixing
\cite {dumbrajs,rijken,hohler1} has not been taken into
account in the present investigation.

The Lagrangian densities corresponding to the interaction
for the vector meson, ${\cal L}_{vec}$, the meson-meson interaction
${\cal L}_0$ and that corresponding to the explicit chiral symmetry breaking
${\cal L}_{SB}$ have been described in detail in references
\cite{paper3,isoamss}.

To investigate the hadronic properties in the medium, we write
the Lagrangian density within the chiral SU(3) model in the mean
field approximation and determine the expectation values
of the meson fields by minimizing the thermodynamical potential
\cite{hartree,kristof1}.

\section{Kaon (antikaon) interactions in the chiral SU(3) model}
\label{kmeson}

In this section, we derive the dispersion relations
for the $K (\bar K)$ \cite{kmeson} and calculate their optical potentials
in asymmetric \cite{isoamss,isoamss1} hadronic matter, taking into the
effects from hyperons. The modifications of the kaon and antikaon energies
arise due to the effects from scalar mesons, scalar-isovector $\delta$ meson
and due to interactions with the nucleons and hyperons. The interactions
of kaons and antikaons with the baryon octet is due to the vectorial
Weinbberg-Tomazawa as well as due to terms similar to the range term
in chiral perturbation theory. 
It might be noted here that the interaction of the pseudoscalar
mesons to the vector mesons, in addition to the pseudoscalar meson--nucleon
vectorial interaction leads to a double counting in the linear realization
of the chiral effective theory \cite{borasoy}. Within the
nonlinear realization of the chiral effective theories, such an interaction
does not arise in the leading or sub-leading order, but only as a higher
order contribution \cite{borasoy}. Hence the vector meson-pesudoscalar
interaction will not be considered within the present investigation.
In the following, we shall derive the dispersion relations for the kaons
and antikaons and study the dependence of the kaon
and antikaon optical potentials on the isospin asymmetric parameter,
$\eta = \frac {1}{2} (\rho_n-\rho_p)/\rho_B$. For this, we shall include 
the effects from isospin asymmetry originating from the scalar-isovector 
$\delta$ field, vectorial interaction with the nucleons,
an isospin symmetric range term \cite{isoamss} as well as an isospin
dependent range term arising from the interaction with the nucleons
\cite{isoamss1}. In the present investigation, we include the effects
from the hypoerons to study the kaon and antikaon properties,
 which were not taken into account in the earlier works.

\begin{figure}
%\vspace{-0.4cm}
%\begin{center}
%\begin{tabular}{c c }
\includegraphics[width=16cm,height=16cm]{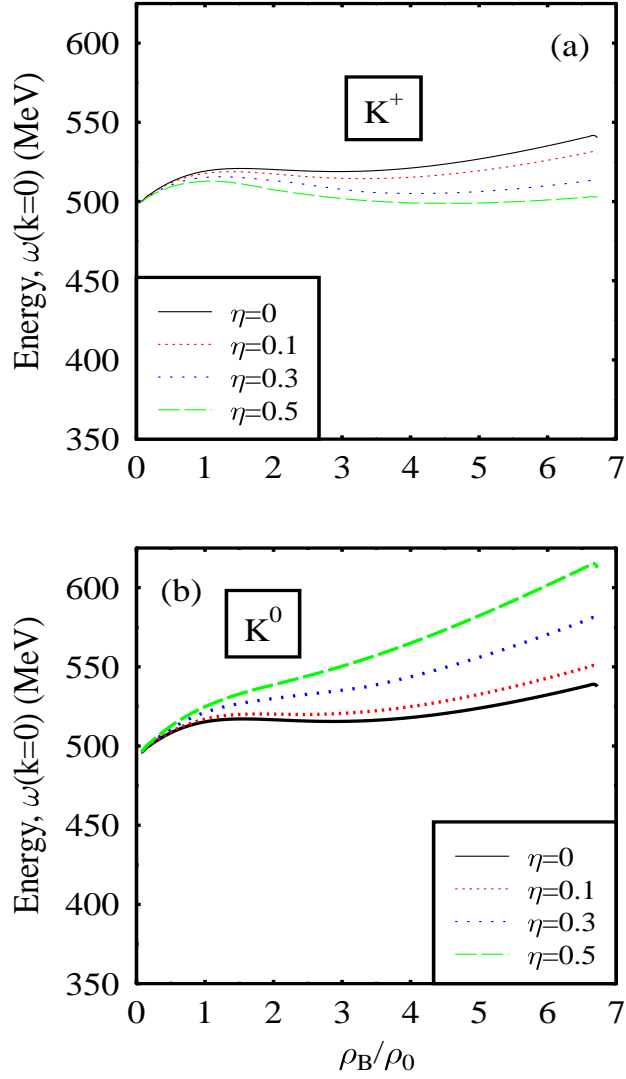}
\caption{
The kaon energies (for $K^+$ in (a) and for $K^0$ in (b)) in MeV plotted
as a functions of the baryon density, $\rho_B/\rho_0$
for $f_s$=0.1 and for different values of the isospin asymmetry  parameter,
$\eta =\frac {1}{2} (\rho_n-\rho_p)/\rho_B$.
}
\end{figure}

\begin{figure}
%\vspace{-0.4cm}
%\begin{center}
%\begin{tabular}{c c }
\includegraphics[width=16cm,height=16cm]{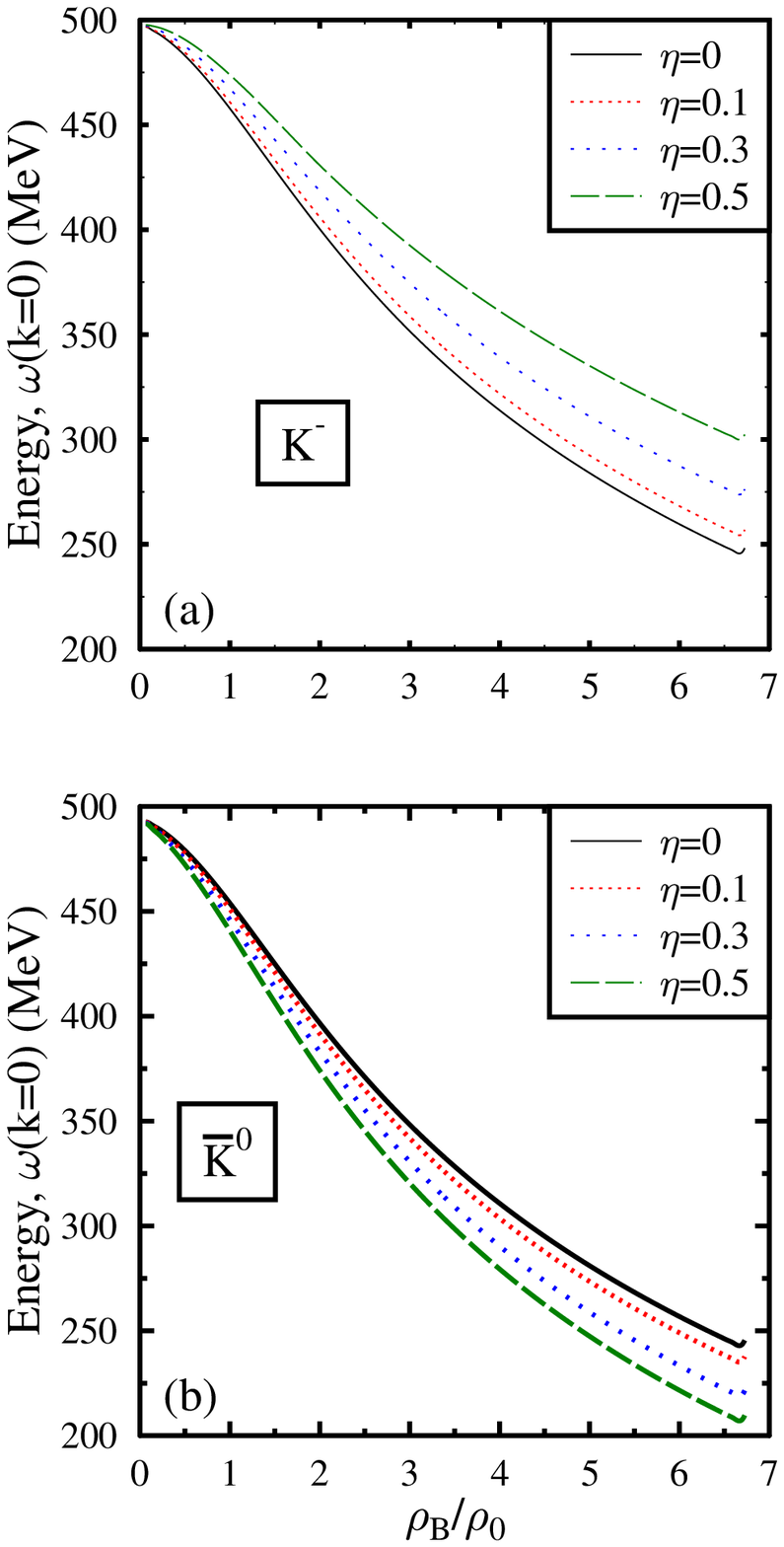}
\caption{
The energies of the antikaons (for $K^-$ in (a) and for $\bar {K^0}$ in (b)),
at zero momentum as functions of the baryon density
($\rho_B/\rho_0$),
are plotted for $f_s$=0.1 and for
 different values of the isospin asymmetry parameter, $\eta$.
}
\end{figure}

\begin{figure}
%\vspace{-0.4cm}
%\begin{center}
%\begin{tabular}{c c }
\includegraphics[width=16cm,height=16cm]{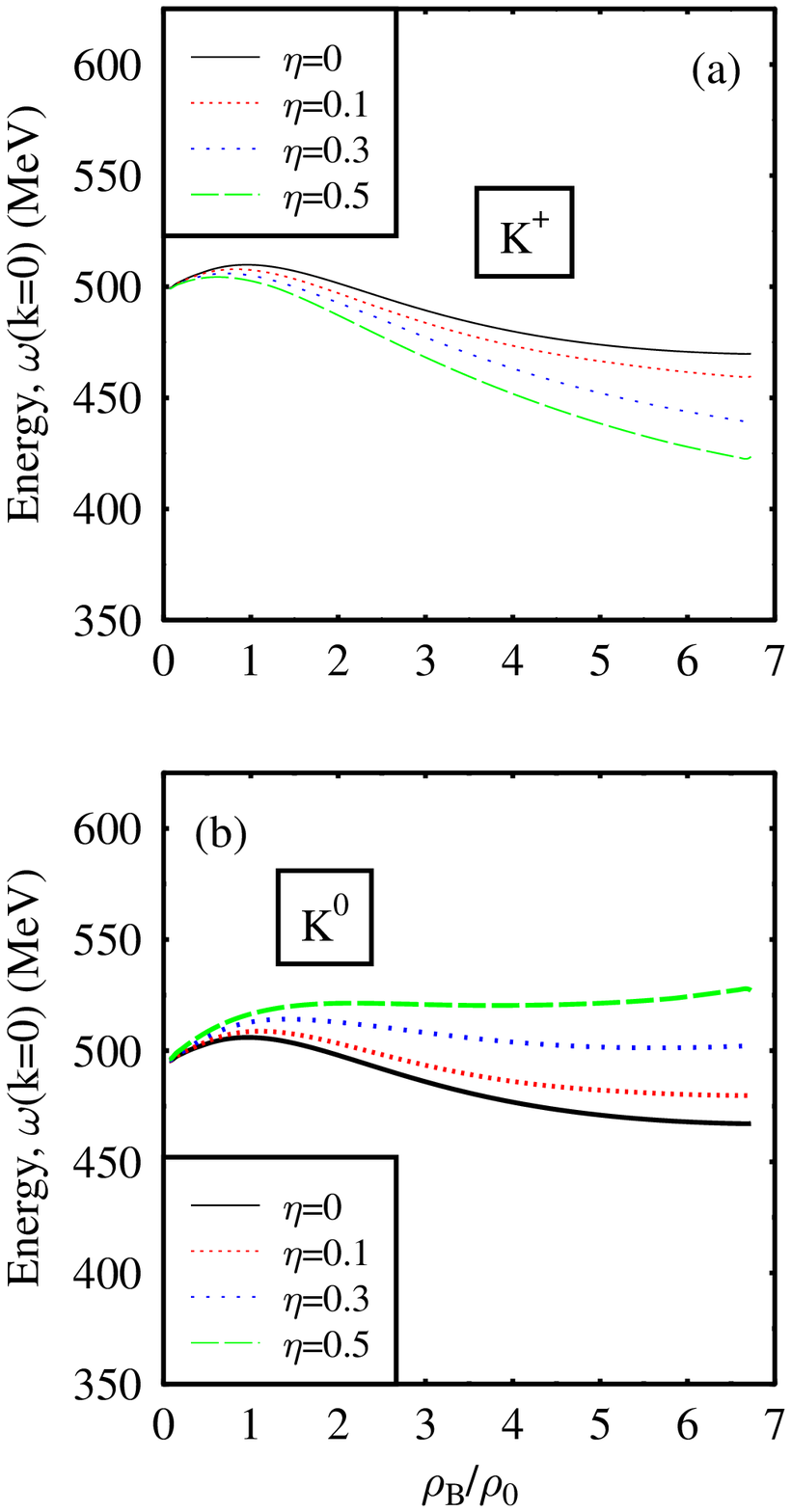}
\caption{
The kaon energies (for $K^+$ in (a) and for $K^0$ in (b)) in MeV plotted
as a functions of the baryon density, $\rho_B/\rho_0$
for $f_s$=0.3 and for different values of the isospin asymmetry  parameter,
$\eta =\frac {1}{2} (\rho_n-\rho_p)/\rho_B$.
}
\end{figure}

\begin{figure}
%\vspace{-0.4cm}
%\begin{center}
%\begin{tabular}{c c }
\includegraphics[width=16cm,height=16cm]{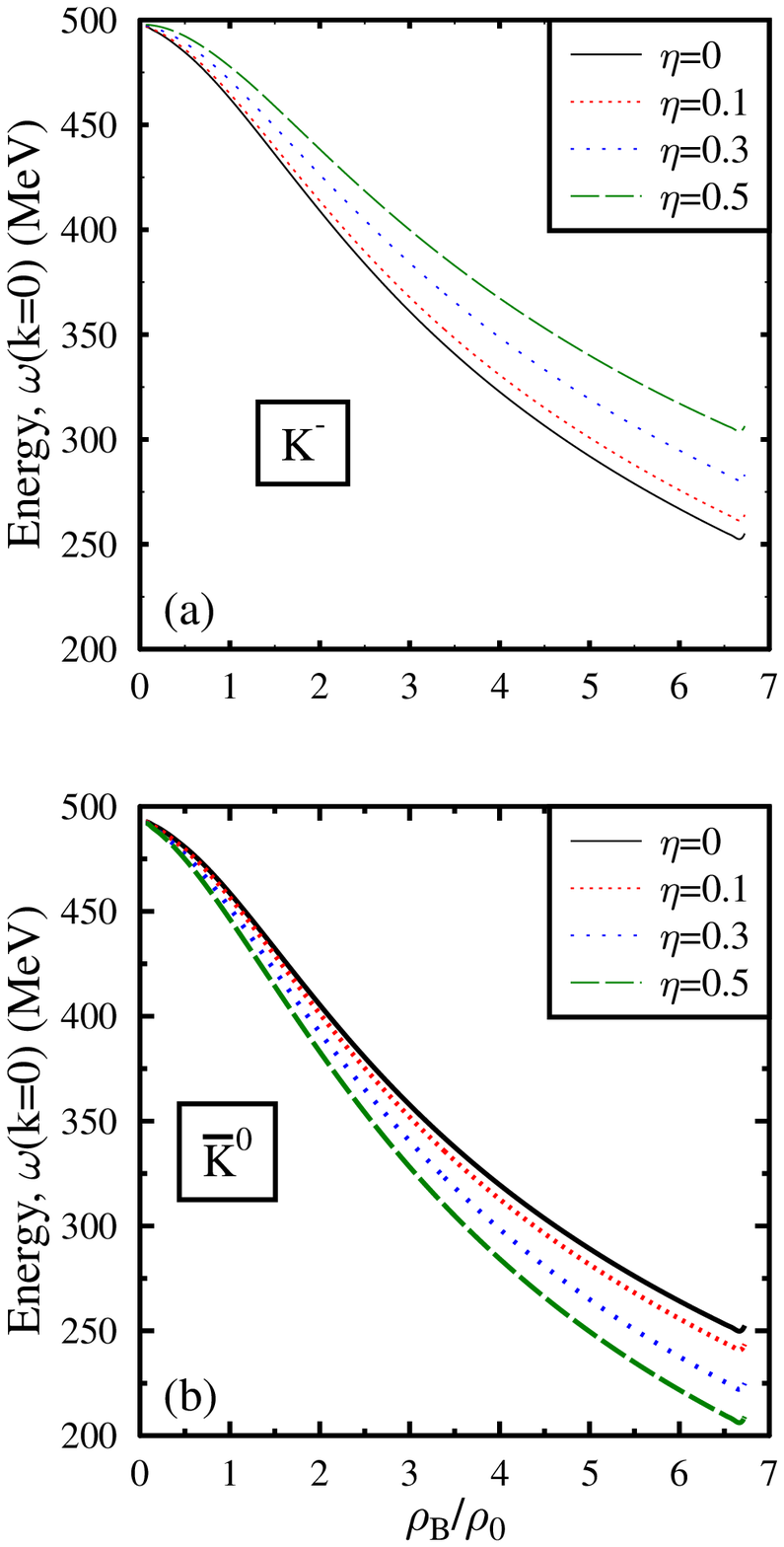}
\caption{
The energies of the antikaons (for $K^-$ in (a) and for $\bar {K^0}$ in (b)),
at zero momentum as functions of the baryon density
($\rho_B/\rho_0$),
are plotted for $f_s$=0.3 and for
 different values of the isospin asymmetry parameter, $\eta$.
}
\end{figure}

\begin{figure}
%\vspace{-0.4cm}
%\begin{center}
%\begin{tabular}{c c }
\includegraphics[width=16cm,height=16cm]{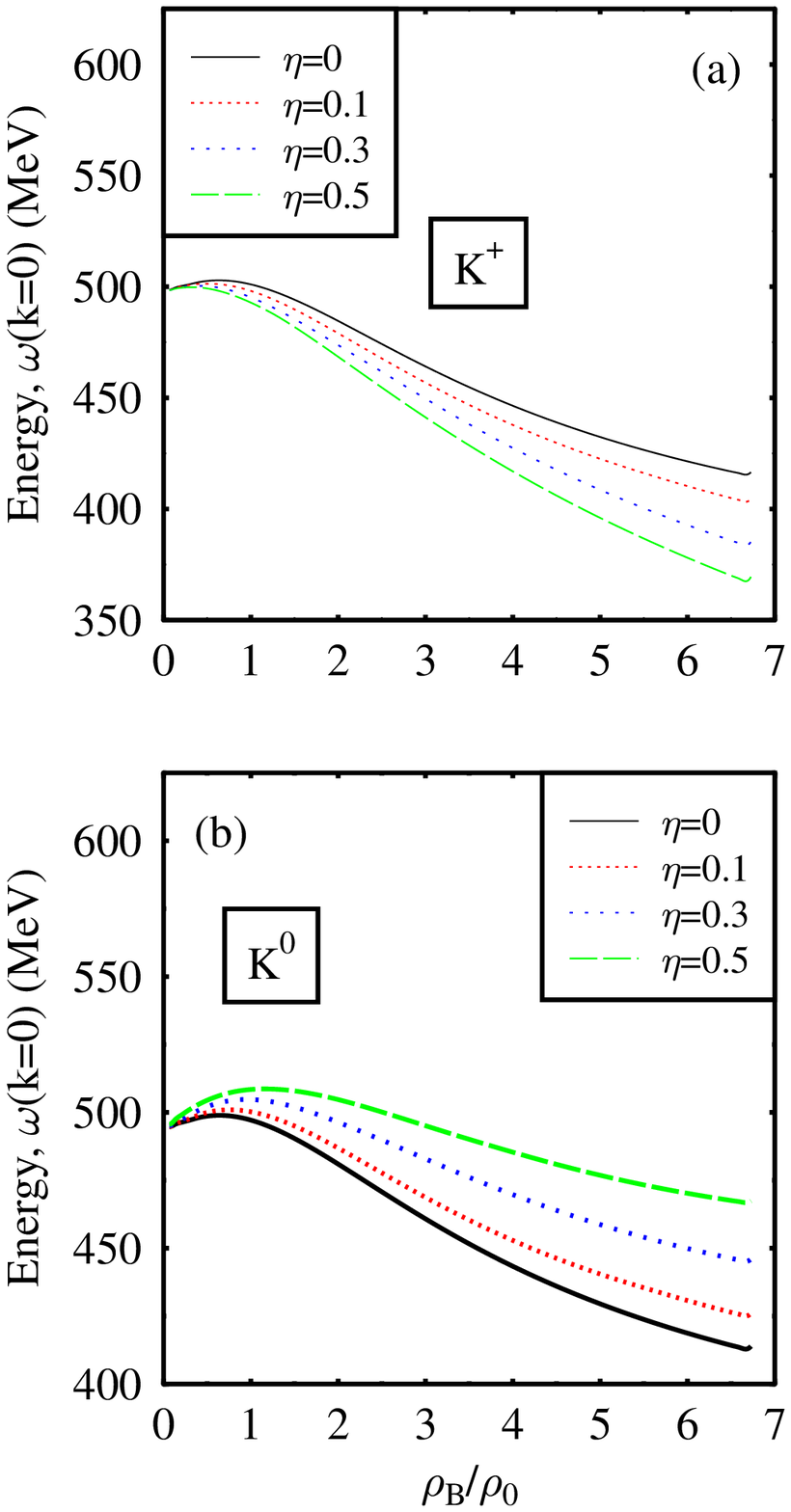}
\caption{
The kaon energies (for $K^+$ in (a) and for $K^0$ in (b)) in MeV plotted
as a functions of the baryon density, $\rho_B/\rho_0$
for $f_s$=0.5 and for different values of the isospin asymmetry  parameter,
$\eta =\frac {1}{2} (\rho_n-\rho_p)/\rho_B$.
}
\end{figure}

\begin{figure}
%\vspace{-0.4cm}
%\begin{center}
%\begin{tabular}{c c }
\includegraphics[width=16cm,height=16cm]{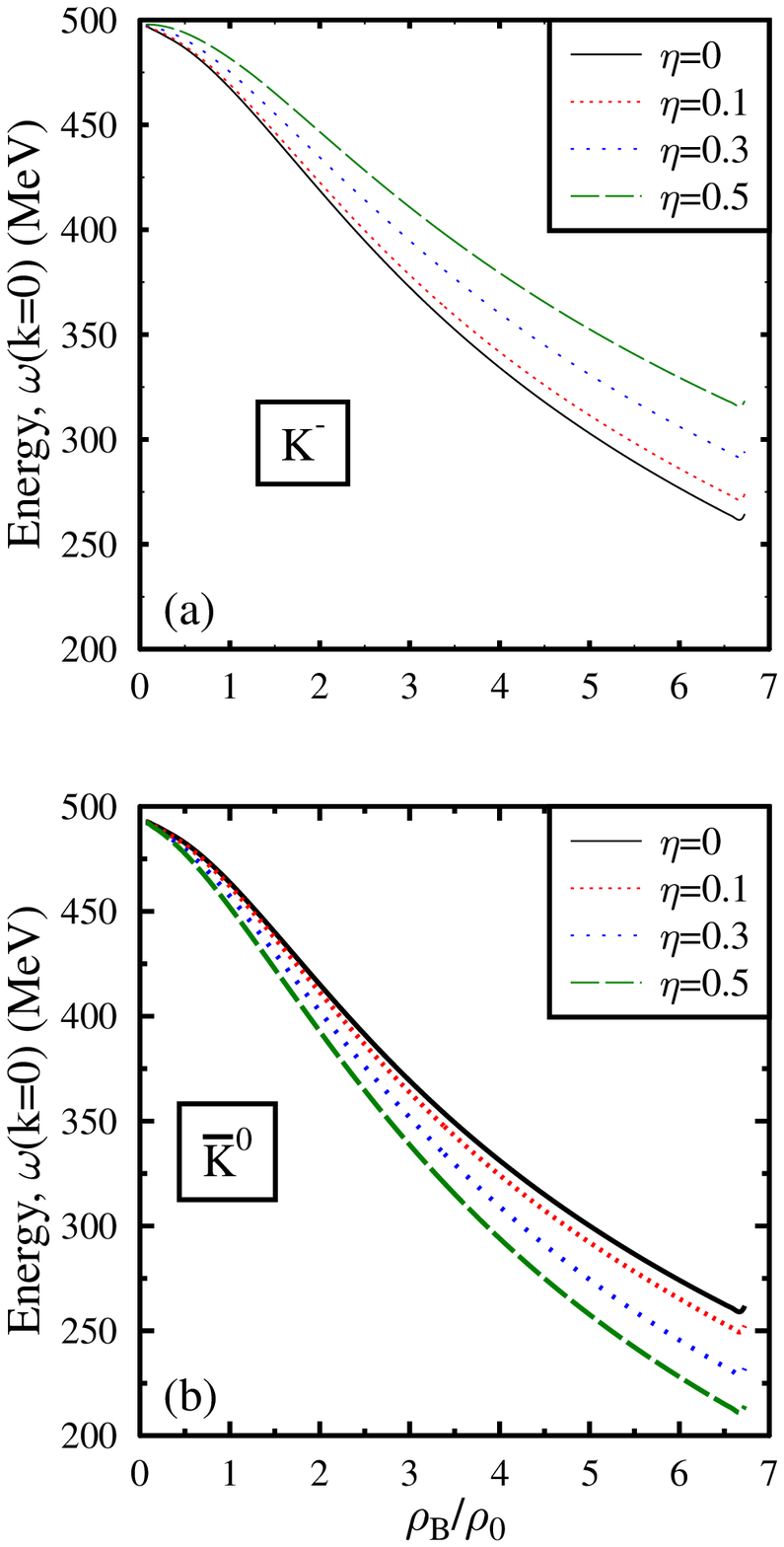}
\caption{
The energies of the antikaons (for $K^-$ in (a) and for $\bar {K^0}$ in (b)),
at zero momentum as functions of the baryon density
($\rho_B/\rho_0$),
are plotted for $f_s$=0.5 and for
 different values of the isospin asymmetry parameter, $\eta$.
}
\end{figure}

\begin{figure}
%\vspace{-0.4cm}
%\begin{center}
%\begin{tabular}{c c }
\includegraphics[width=20cm,height=20cm]{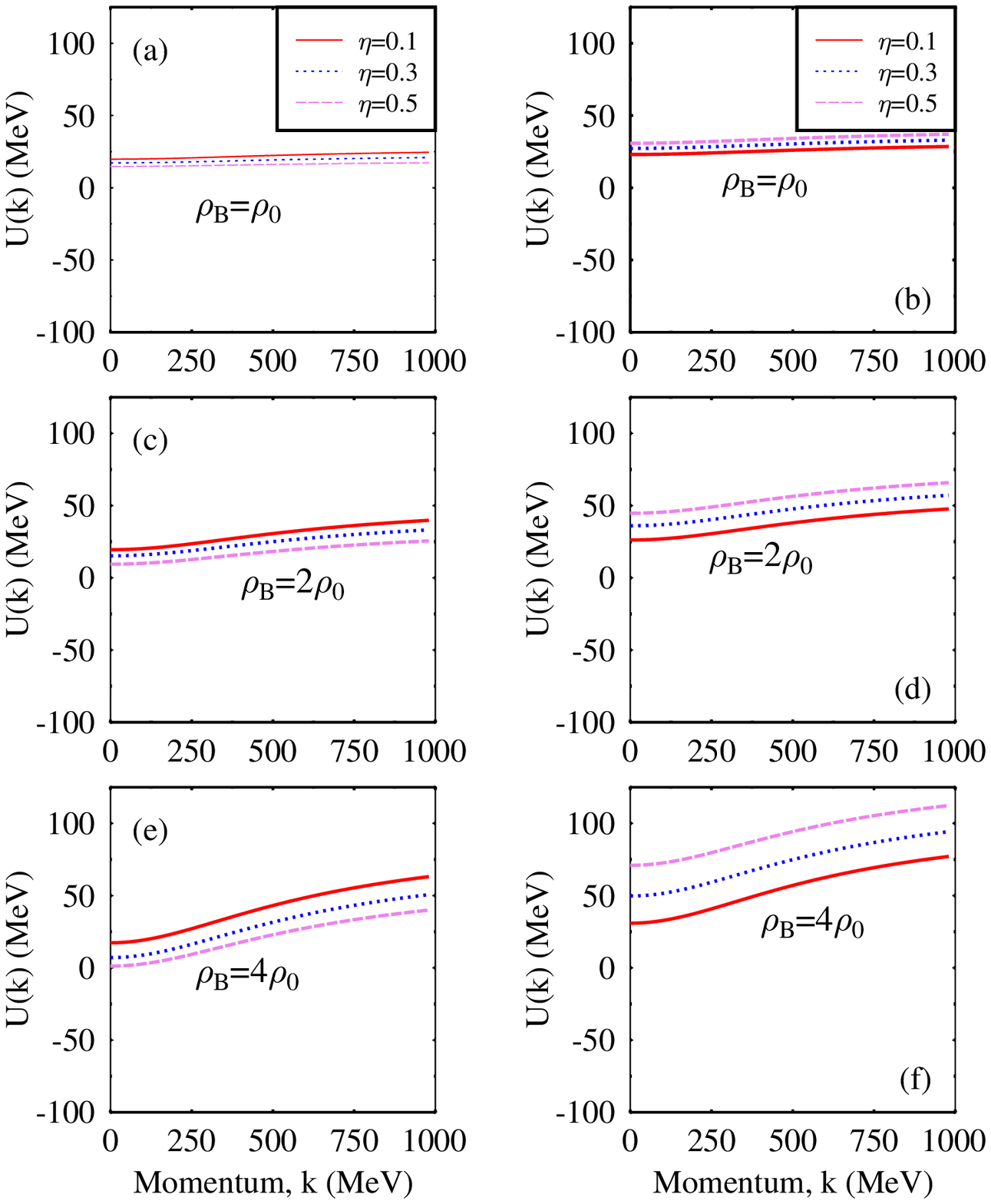}
\caption{
The kaon optical potentials (for $K^+$ in (a), (c) and (e) and for $K^0$
in (b), (d) and (f)) in MeV for $f_s$=0.1,
plotted as functions of the momentum for various baryon densities, $\rho_B$
and for different values of the isospin asymmetry  parameter, $\eta$.
}
\end{figure}

\begin{figure}
%\vspace{-0.4cm}
%\begin{center}
%\begin{tabular}{c c }
\includegraphics[width=20cm,height=20cm]{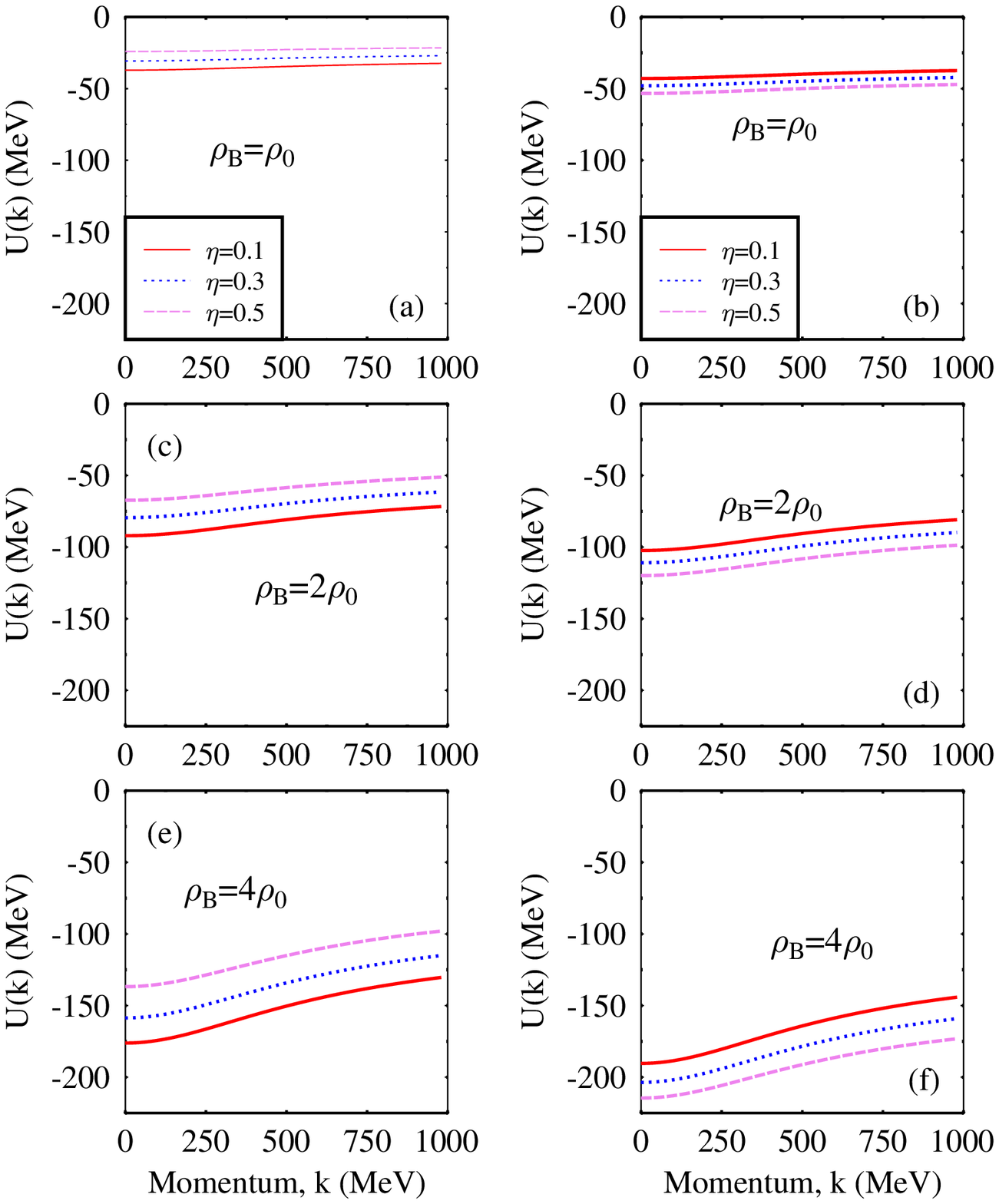}
\caption{
The antikaon optical potentials (for $K^-$ in (a), (c) and (e) and for
$\bar {K^0}$ in (b), (d) and (f)) in MeV for $f_s$=0.1,
plotted as functions of the momentum for various baryon densities, $\rho_B$
and for different values of the isospin asymmetry  parameter, $\eta$.
}
\end{figure}

\begin{figure}
%\vspace{-0.4cm}
%\begin{center}
%\begin{tabular}{c c }
\includegraphics[width=20cm,height=20cm]{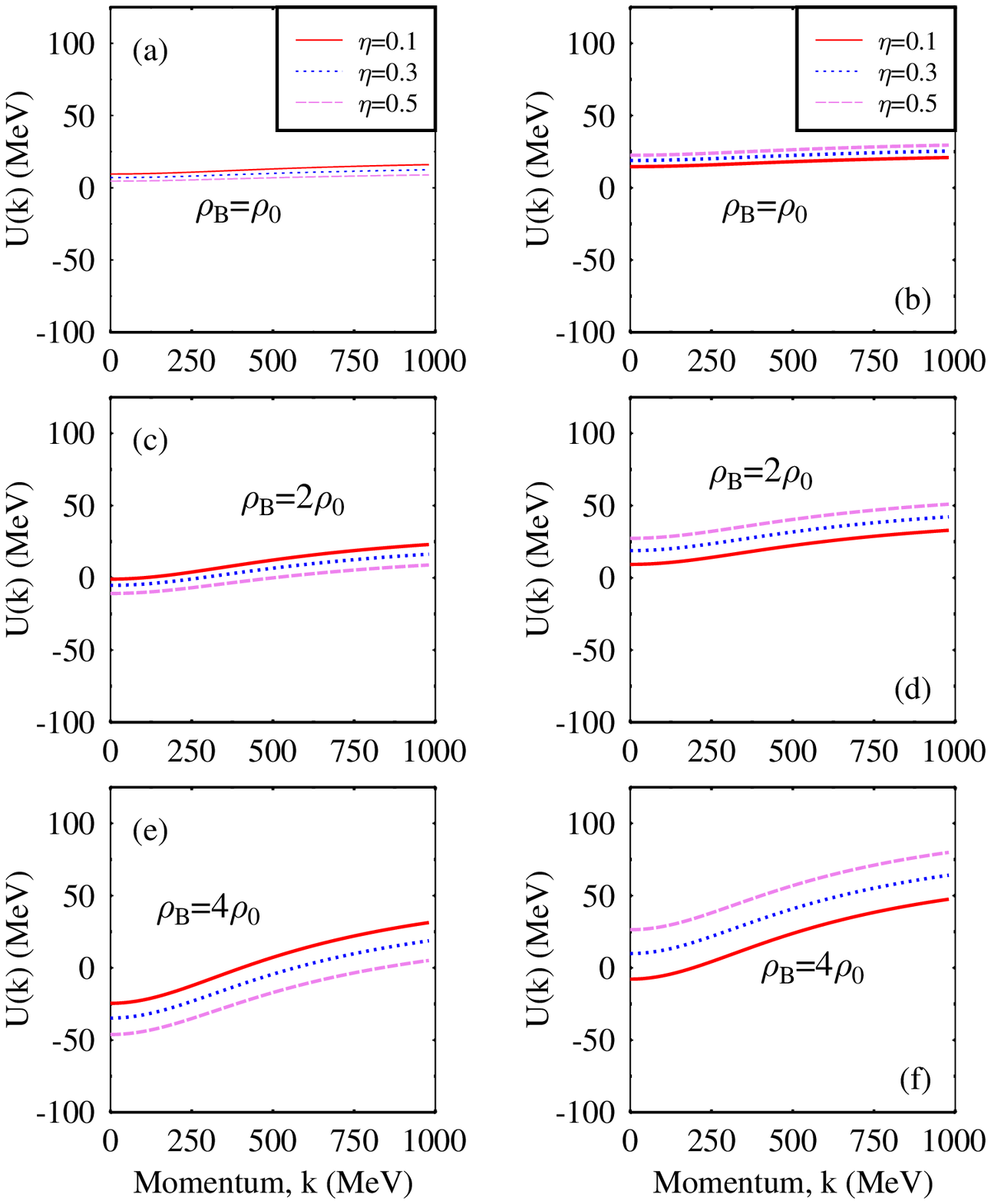}
\caption{
The kaon optical potentials (for $K^+$ in (a), (c) and (e) and for $K^0$
in (b), (d) and (f)) in MeV for $f_s$=0.3,
plotted as functions of the momentum for various baryon densities, $\rho_B$
and for different values of the isospin asymmetry  parameter, $\eta$.
}
\end{figure}

\begin{figure}
%\vspace{-0.4cm}
%\begin{center}
%\begin{tabular}{c c }
\includegraphics[width=20cm,height=20cm]{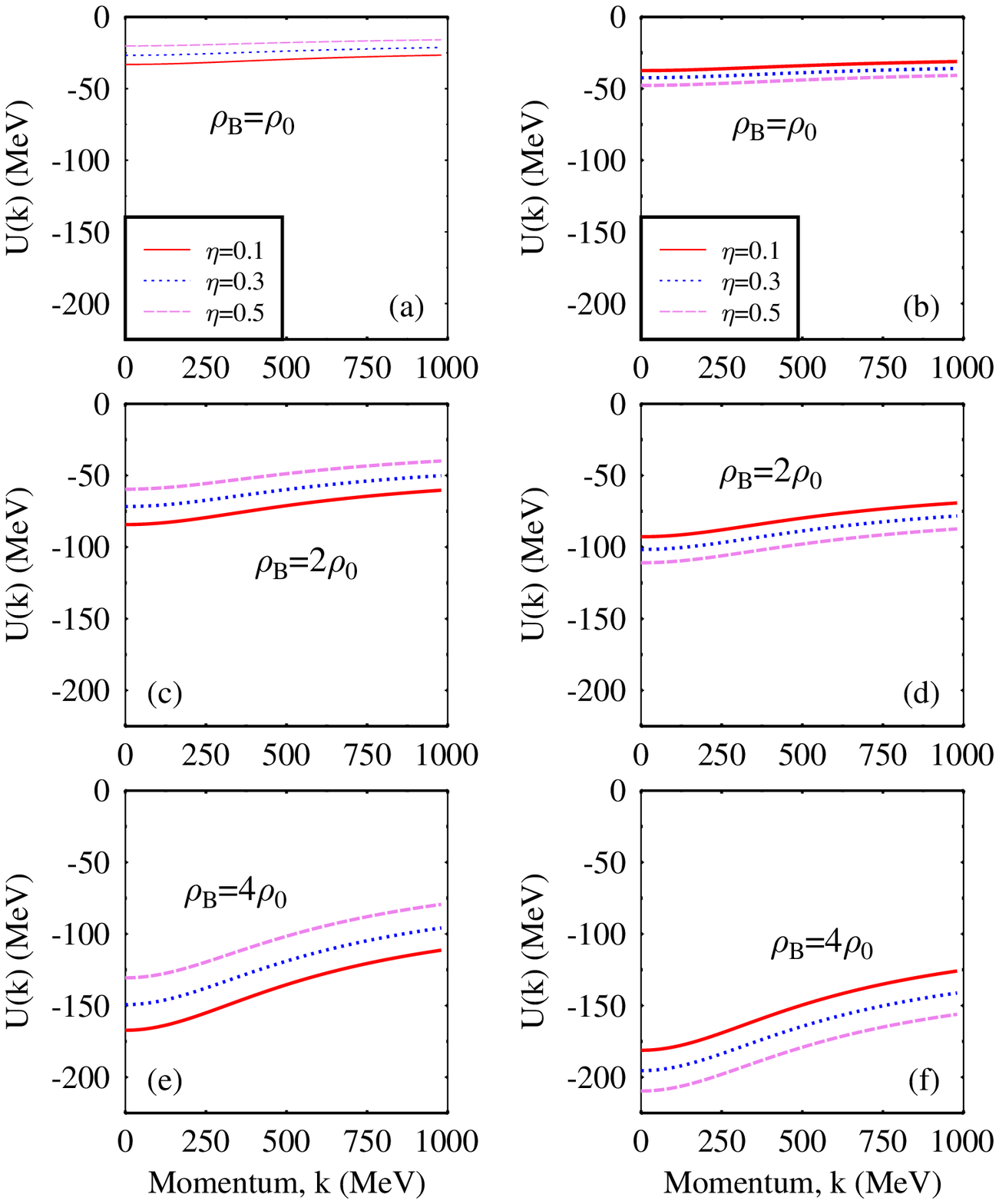}
\caption{
The antikaon optical potentials (for $K^-$ in (a), (c) and (e) and for
$\bar {K^0}$ in (b), (d) and (f)) in MeV for $f_s$=0.3,
plotted as functions of the momentum for various baryon densities, $\rho_B$
and for different values of the isospin asymmetry  parameter, $\eta$.
}
\end{figure}

\begin{figure}
%\vspace{-0.4cm}
%\begin{center}
%\begin{tabular}{c c }
\includegraphics[width=20cm,height=20cm]{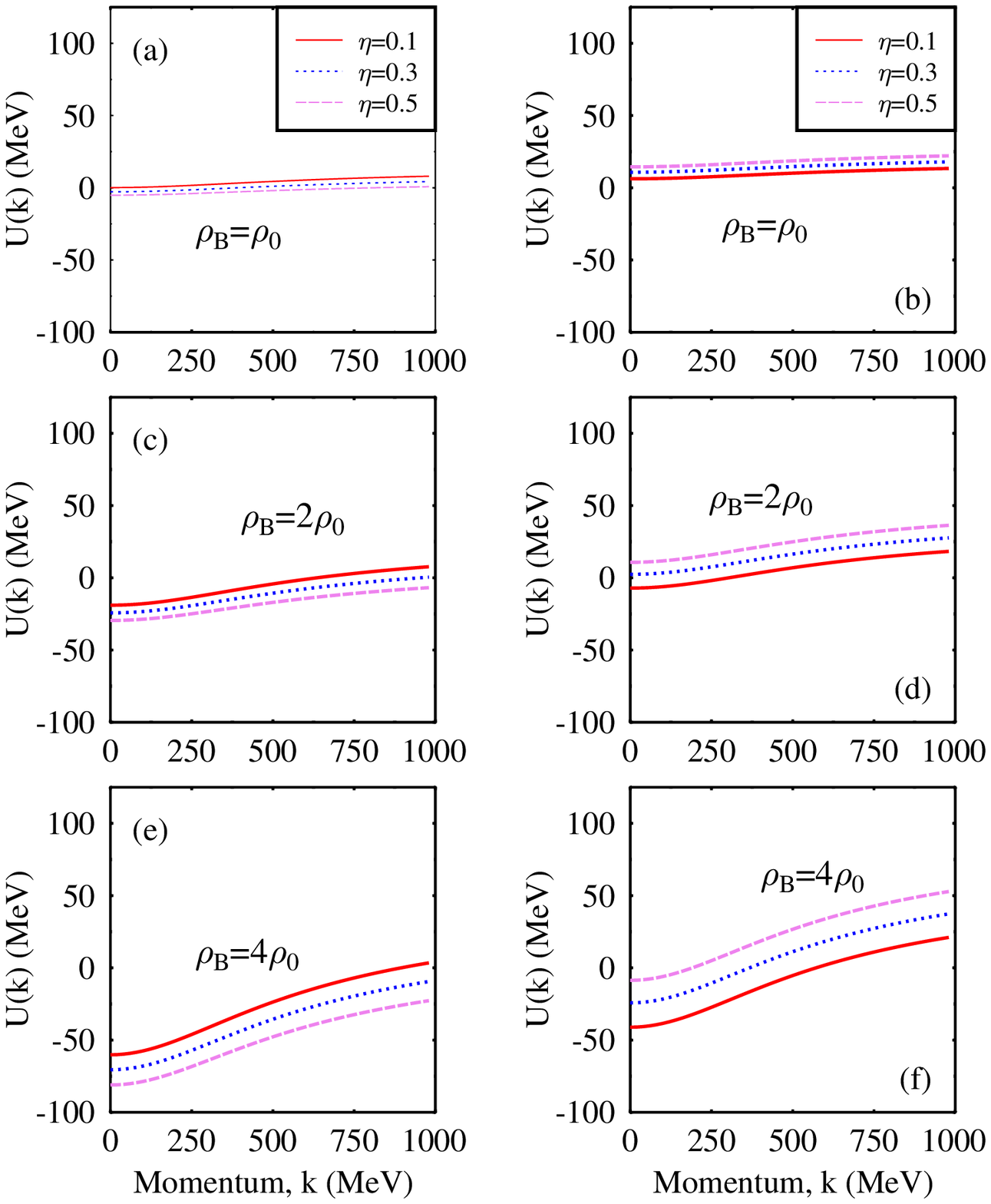}
\caption{
The kaon optical potentials (for $K^+$ in (a), (c) and (e) and for $K^0$
in (b), (d) and (f)) in MeV for $f_s$=0.5,
plotted as functions of the momentum for various baryon densities, $\rho_B$
and for different values of the isospin asymmetry  parameter, $\eta$.
}
\end{figure}

\begin{figure}
%\vspace{-0.4cm}
%\begin{center}
%\begin{tabular}{c c }
\includegraphics[width=20cm,height=20cm]{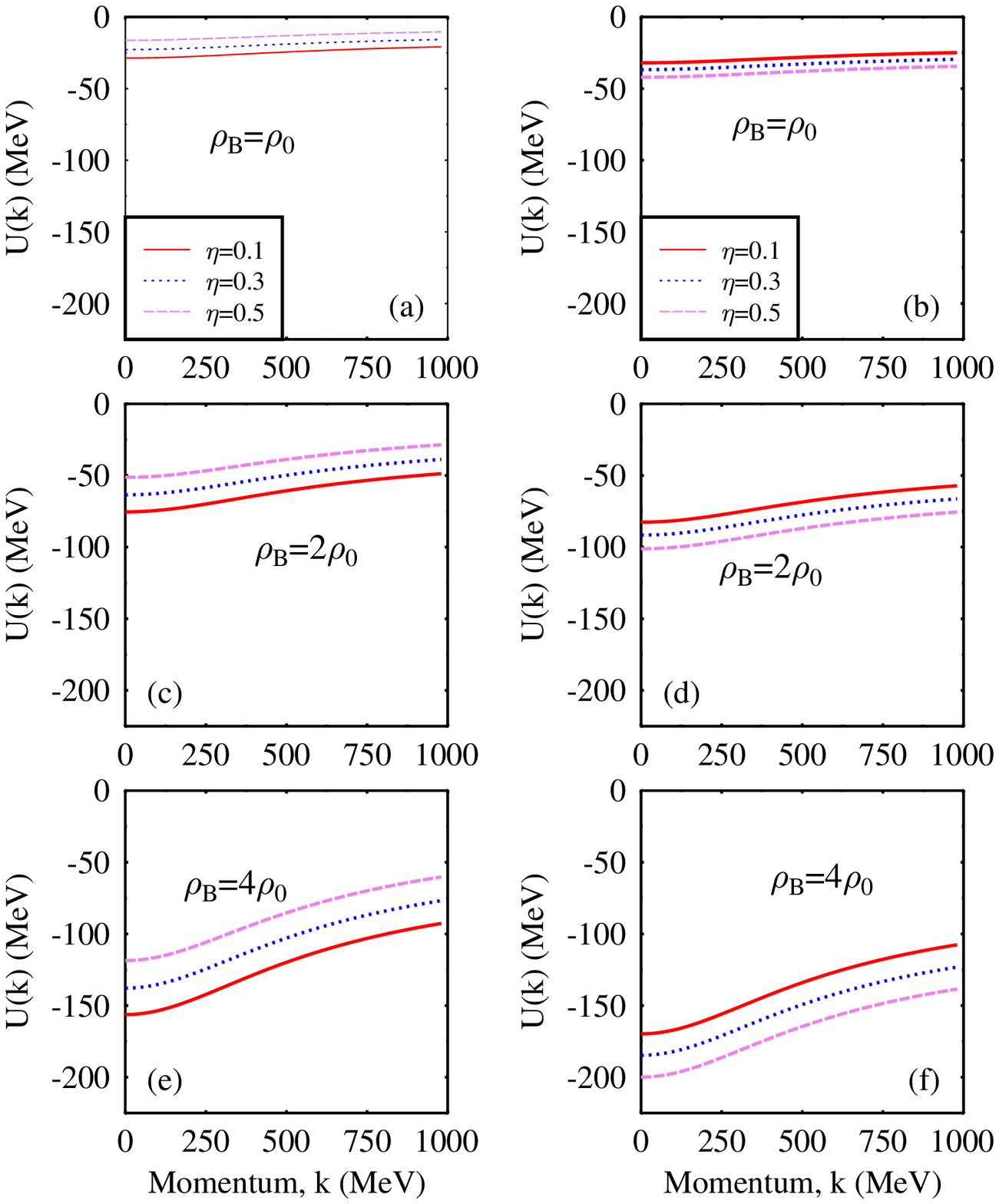}
\caption{
The antikaon optical potentials (for $K^-$ in (a), (c) and (e) and for
$\bar {K^0}$ in (b), (d) and (f)) in MeV for $f_s$=0.5,
plotted as functions of the momentum for various baryon densities, $\rho_B$
and for different values of the isospin asymmetry  parameter, $\eta$.
}
\end{figure}

The interaction Lagrangian modifying the energies of the $K(\bar K)$-mesons
is given as
\begin{eqnarray}
\cal L _{KN} & = & -\frac {i}{4 f_K^2} \Big [\Big ( 2 \bar p \gamma^\mu p
+\bar n \gamma ^\mu n -\bar {\Sigma^-}\gamma ^\mu \Sigma ^-
+\bar {\Sigma^+}\gamma ^\mu \Sigma ^+
- 2\bar {\Xi^-}\gamma ^\mu \Xi ^-
- \bar {\Xi^0}\gamma ^\mu \Xi^0 \Big)
\nonumber \\
& \times &
\Big(K^- (\partial_\mu K^+) - (\partial_\mu {K^-})  K^+ \Big )
\nonumber \\
& + &
\Big ( \bar p \gamma^\mu p
+ 2\bar n \gamma ^\mu n +\bar {\Sigma^-}\gamma ^\mu \Sigma ^-
-\bar {\Sigma^+}\gamma ^\mu \Sigma ^+
- \bar {\Xi^-}\gamma ^\mu \Xi ^-
- 2 \bar {\Xi^0}\gamma ^\mu \Xi^0 \Big)
\nonumber \\
& \times &
\Big(\bar {K^0} (\partial_\mu K^0) - (\partial_\mu {\bar {K^0}})  K^0 \Big )
\Big ]
\nonumber \\
 &+ & \frac{m_K^2}{2f_K} \Big [ (\sigma +\sqrt 2 \zeta+\delta)(K^+ K^-)
 + (\sigma +\sqrt 2 \zeta-\delta)(K^0 \bar { K^0})
\Big ] \nonumber \\
& - & \frac {1}{f_K}\Big [ (\sigma +\sqrt 2 \zeta +\delta)
(\partial _\mu {K^+})(\partial ^\mu {K^-})
+(\sigma +\sqrt 2 \zeta -\delta)
(\partial _\mu {K^0})(\partial ^\mu \bar {K^0})
\Big ]
\nonumber \\
&+ & \frac {d_1}{2 f_K^2}(\bar p p +\bar n n +\bar {\Lambda^0}{\Lambda^0}
+\bar {\Sigma ^+}{\Sigma ^+}
+\bar {\Sigma ^0}{\Sigma ^0}
+\bar {\Sigma ^-}{\Sigma ^-}
+\bar {\Xi ^-}{\Xi ^-}
+\bar {\Xi ^0}{\Xi ^0}
 )\nonumber \\
&\times & \big ( (\partial _\mu {K^+})(\partial ^\mu {K^-})
+(\partial _\mu {K^0})(\partial ^\mu {\bar {K^0}})
\big )
\nonumber \\
&+& \frac {d_2}{2 f_K^2} \Big [
(\bar p p+\frac {5}{6} \bar {\Lambda^0}{\Lambda^0}
+\frac {1}{2} \bar {\Sigma^0}{\Sigma^0}
+\bar {\Sigma^+}{\Sigma^+}
+\bar {\Xi^-}{\Xi^-}
+\bar {\Xi^0}{\Xi^0}
) (\partial_\mu K^+)(\partial^\mu K^-) 
\nonumber \\
 &+ &(\bar n n
+\frac {5}{6} \bar {\Lambda^0}{\Lambda^0}
+\frac {1}{2} \bar {\Sigma^0}{\Sigma^0}
+\bar {\Sigma^-}{\Sigma^-}
+\bar {\Xi^-}{\Xi^-}
+\bar {\Xi^0}{\Xi^0}
) (\partial_\mu K^0)(\partial^\mu {\bar {K^0}})
\Big ]
\label{lagd}
\end{eqnarray}
In (\ref{lagd}) the first term is the vectorial interaction term
(Weinberg-Tomozawa term) obtained from the kinetic term of the Lagrangian,
the second and third terms are obtained from the explicit symmetry breaking 
and the pseudoscalar kinetic terms of the chiral effective Lagrangian 
repectively \cite{isoamss,isoamss1}. The fourth and fifth terms 
in (\ref{lagd}) for the KN interactions arise from the terms
\begin{equation}
{\cal L }_{(d_1)}^{BM} =\frac {d_1}{2} Tr (u_\mu u ^\mu)Tr( \bar B B),
\end{equation}
and,
\begin{equation}
{\cal L }_{(d_2)}^{BM} =d_2 Tr (\bar B u_\mu u ^\mu B).
\end{equation}
in the SU(3) chiral model \cite{kmeson1,isoamss}. The last three
terms in (\ref{lagd}) represent the range term in the chiral
model, with the last term being an isospin asymmetric interaction.
We might note here that in equation (\ref{lagd}), we have not
written the terms which are of the form $\bar B_i B_j$, with $i \ne j$.
These types of terms are however relevant for the calculation
of the kaon-nucloen scattering terms \cite{isoamss1}.
The Fourier transformation of the equation-of-motion for kaons
(antikaons) leads to the dispersion relations,
$$-\omega^2+ {\vec k}^2 + m_K^2 -\Pi(\omega, |\vec k|)=0,$$
where $\Pi$ denotes the kaon (antikaon) self energy in the medium.

Explicitly, the self energy $\Pi (\omega,|\vec k|)$ for the kaon doublet,
($K^+$,$K^0$) 
arising from the interaction (\ref{lagd}) is given as
\begin{eqnarray}
\Pi (\omega, |\vec k|) &= & -\frac {1}{4 f_K^2}\Big [3 (\rho_p +\rho_n)
\pm (\rho_p -\rho_n) \pm 2 (\rho_{\Sigma^+}-\rho_{\Sigma^-})
-\big ( 3 (\rho_{\Xi^-} +\rho_{\Xi^0}) \pm (\rho_{\Xi^-} -\rho_{\Xi^0})
\big)
\Big ] \omega\nonumber \\
&+&\frac {m_K^2}{2 f_K} (\sigma ' +\sqrt 2 \zeta ' \pm \delta ')
\nonumber \\ & +& \Big [- \frac {1}{f_K}
(\sigma ' +\sqrt 2 \zeta ' \pm \delta ')
+\frac {d_1}{2 f_K ^2} (\rho_s ^p +\rho_s ^n
+{\rho^s} _{\Lambda^0}+{\rho^s} _{\Sigma^+}+{\rho^s} _{\Sigma^0}
+{\rho^s} _{\Sigma^-} +{\rho^s} _{\Xi^-} +{\rho^s} _{\Xi^0}
)\nonumber \\
&+&\frac {d_2}{4 f_K ^2} \Big (({\rho^s} _p +{\rho^s} _n)
\pm   ({\rho^s} _p -{\rho^s} _n)
+{\rho^s} _{\Sigma ^0}+\frac {5}{3} {\rho^s} _{\Lambda^0}
+ ({\rho^s} _{\Sigma ^+}+{\rho^s} _{\Sigma ^-})
\pm ({\rho^s} _{\Sigma ^+}-{\rho^s} _{\Sigma ^-})\nonumber \\
 &+ & 2 {\rho^s} _ {\Xi^-}+
2 {\rho^s} _ {\Xi^0}
\Big )
\Big ]
(\omega ^2 - {\vec k}^2),
\end{eqnarray}
where the $\pm$ signs refer to the $K^+$ and $K^0$ respectively.
In the above, $\sigma'(=\sigma-\sigma _0)$,
$\zeta'(=\zeta-\zeta_0)$ and  $\delta'(=\delta-\delta_0)$
are the fluctuations of the scalar-isoscalar fields $\sigma$ and $\zeta$,
and the third component of the scalar-isovector field, $\delta$,
from their vacuum expectation values.
The vacuum expectation value of $\delta$ is zero ($\delta_0$=0), since
a nonzero value for it will break the isospin symmetry of the vacuum
(the small isospin breaking effect coming from the mass and charge
 difference of the up and down quarks
has been neglected here).
$\rho_i$ and ${\rho^s}_{i}$ with $i=p,n, \Lambda, \Sigma ^\pm,
\Sigma ^0, \Xi ^-, \Xi ^0$
are the number density and the scalar density
of the baryon of type $i$.

Similarly, for the antikaon doublet, ($K^-$,$\bar {K^0}$),
the self-energy is calculated as
\begin{eqnarray}
\Pi (\omega, |\vec k|) &= & \frac {1}{4 f_K^2}\Big [3 (\rho_p +\rho_n)
\pm (\rho_p -\rho_n) \pm 2 (\rho_{\Sigma^+}-\rho_{\Sigma^-})
- \big ( 3 (\rho_{\Xi^-} +\rho_{\Xi^0}) \pm (\rho_{\Xi^-} -\rho_{\Xi^0})
\big)
\Big ] \omega\nonumber \\
&+&\frac {m_K^2}{2 f_K} (\sigma ' +\sqrt 2 \zeta ' \pm \delta ')
\nonumber \\ & +& \Big [- \frac {1}{f_K}
(\sigma ' +\sqrt 2 \zeta ' \pm \delta ')
+\frac {d_1}{2 f_K ^2} (\rho_s ^p +\rho_s ^n
+{\rho^s} _{\Lambda^0}+{\rho^s} _{\Sigma^+}+{\rho^s} _{\Sigma^0}
+{\rho^s} _{\Sigma^-} +{\rho^s} _{\Xi^-} +{\rho^s} _{\Xi^0}
)\nonumber \\
&+&\frac {d_2}{4 f_K ^2} \Big (({\rho^s} _p +{\rho^s} _n)
\pm   ({\rho^s} _p -{\rho^s} _n)
+{\rho^s} _{\Sigma ^0}+\frac {5}{3} {\rho^s} _{\Lambda^0}
+ ({\rho^s} _{\Sigma ^+}+{\rho^s} _{\Sigma ^-})
\pm ({\rho^s} _{\Sigma ^+}-{\rho^s} _{\Sigma ^-})\nonumber \\
 &+ & 2 {\rho^s} _ {\Xi^-}+ 2 {\rho^s} _ {\Xi^0}
\Big )
\Big ]
(\omega ^2 - {\vec k}^2),
\end{eqnarray}
where the $\pm$ signs refer to the $K^-$ and $\bar {K^0}$ respectively.

The optical potentials
are calculated from the energies of the kaons and antikaons using
\be
U(\omega, k) = \omega (k) -\sqrt {k^2 + m_K ^2},
\ee
where $m_K$ is the vacuum mass for the kaon (antikaon).

The parameters $d_1$ and $d_2$ are calculated from the
empirical values of the KN scattering lengths
for I=0 and I=1 channels \cite{isoamss}, which are taken as
\cite{thorsson,juergen,barnes}
\be
a _{KN} (I=0) \approx -0.09 ~ {\rm {fm}},\;\;\;\;
a _{KN} (I=1) \approx -0.31 ~ {\rm {fm}}.
\label{akn01emp}
\ee
leading to the isospin averaged KN scattering length as
\be
\bar a _{KN}=\frac {1}{4} a_{KN}(I=0)+
\frac {3}{4} a_{KN}(I=1) \approx -0.255 ~ \rm {fm}.
\ee

\section{Results and Discussions}
\label{kmass}

The present calculations use the following model parameters. The
values, $g_{\sigma N}=10.6,\;\; {\rm and}\;\; g_{\zeta N}=-0.47$
are determined by fitting vacuum baryon masses. The other
parameters as fitted to the asymmetric nuclear matter saturation
properties in the mean field approximation are: $g_{\omega
N}$=13.3, $g_{\rho N}$=5.5, $g_4$=79.7, $g_{N \delta}$=2.5,
$m_\zeta$ =1024.5 MeV, $m_\sigma$= 466.5 MeV and $m_\delta$=899.5
MeV \cite{isoamss1}. 
%
%The value of ${g_{\rho N}}^2/4\pi \approx 2.4$ of the present
%work, may be compared with the value of 2.6 \cite{hohler2} and
%a range of values of 2.1 to 3.4 \cite{mosel} in the literature.
%The value of the $\omega$ meson-nucleon coupling,
%${g_{\omega N}}^2/4\pi \approx 14$
%in the present investigation is the same as in Ref. \cite{dumbrajs},
%whereas this coupling was taken to be around 24 in Ref. \cite{hohler1}.
%%
%
The coefficients $d_1$ and $d_2$, calculated from the
empirical values of the KN scattering lengths for I=0 and I=1
channels (\ref{akn01emp}), are $2.56/{m_K}$ and $0.73/{m_K}$
respectively. 

The kaon and antikaon properties were studied
in the isospin symmetric nuclear matter
within the chiral SU(3) model in ref. \cite{kmeson1}
and in the isospin asymmetric nuclear matter in ref.s
\cite{isoamss,isoamss1}.
In the present work, we investigate also the effects of the hyperons
on the energies of the kaons and antikaons in the strange hadronic matter. 
The contribution from the vector interaction (Weinberg-Tomozawa
term) leads to a drop for the antikaon energy, whereas they
are repulsive for the kaons in the asymmetric nuclear medium. 
There are now contributions from hyperons due to this vectorial
interaction as given in the first term in equation (\ref{lagd}).
One may note that there is no contribution arising from either 
$\Lambda^0$ or the $\Sigma^0$- hyperon, for such a baryon-pesudoscalar 
meson vectorial interaction. The scalar meson
exchange term arising from the scalar-isoscalar fields ($\sigma$
and $\zeta$) is attractive for both $K$ and $\bar K$ doublets. 
The first term of the range term of eq. (\ref{lagd}) is
repulsive, whereas the second term has an isospin symmetric 
attractive contribution for both kaons
and antikaons, with contributions arising due to interactions
with the baryon octet. The third term of the range term has an isospin
dependence \cite{isoamss1} which now takes into account
the effects of hyperons, in addition to contributions from the nucleons.

The isospin asymmetries within the kaon doublet ($K^+$, $K^0$)
as well as in antikaon ($K^-$, $\bar K^0$) energies arise
from the Weinberg-Tomazawa term due to
asymmetry in proton and neutron number densities as well as
asymmetries in the hyperon number densities.
They also arise from the scalar-isovector $\delta$ field
becoming nonzero for the isospin asymmetric hadronic matter
as can be seen from the second term (the scalar exchange term)
as well as the in the range term given by the third term 
in equation (\ref{lagd}).
The $d_2$ term in the interaction Lagrangian given by equation
(\ref{lagd}) also introduces asymmetries for $K^+$ and $K^0$,
as well as for antikaon ($K^-$ and $\bar {K^0}$)
-energies, arising from both the nucleon as well as hyperonic 
sectors.  For $\rho_n > \rho_p$, in the kaon sector,
$K^+$ ($K^0$) has negative (positive) contributions from $\delta$.
The $\delta$ contribution from the scalar exchange term is
positive (negative) for $K^+$ ($K^0$), whereas that arising from
the range term has the opposite sign and dominates over the former
contribution.

In figures 1, 3 and 5, the isospin dependent energies 
of the $K^+$ and $K^0$ at zero momentum,
are plotted as functions of densities 
for various values of the strangeness fraction, 
$f_s=\frac {\sum_{i} \rho_{H_i}}{2\rho_B}$.
For $f_s$=0.1 plotted in figure 1, 
one sees that when the density
is increased, the masses of the kaons initially increase upto
about $\rho_B=\rho_0$. However, as the density is further
increased, the kaon masses do not show a monotonic rise, but a drop
at around this value.  When the strangeness fraction is  increased,
the masses of the kaons are seen to decrease with increasing density,
as one can see from figures 3 and 5, which correspond to the strangeness
fractions, $f_s$=0.3 and 0.5 respectively. This is due to the fact 
that the second term of the range term (the $d_1$ term) which is
attractive, becomes more and more dominant with
increasing contributions from the hyperons, in addition to nucleons.

For $f_s$=0.1, at $\rho_B=\rho_0$, 
the energy of $K^+$ is seen to drop by about 7 MeV 
at zero momentum when $\eta$ changes from 0 to 0.5 whereas
the $K^0$ energy is seen to increase by about 10 MeV 
from the isospin zero case. The reason for the
opposite behavior for the  $K^+$ and $K^0$ on the isospin
asymmetry in the nuclear medium \cite{isoamss,isoamss1}
originates from the vectorial (Weinberg-Tomozawa),
$\delta$ meson contribution as well as from the isospin dependent
range term ($d_2$- term) contributions.
For $K^+$, the $\eta$-dependence of
the energy is seen to be less sensitive at higher densities,
whereas the energy of $K^0$  is seen to have a larger drop from
the $\eta$=0 case, as we increase the density.
The isospin asymmetry in medium modifications of the $K^+$ and $K^0$
mesons are seen to be particularly significant for higher strangeness
fractions at high densities. 
For $f_s$=0.5, at $\rho_B=\rho_0$, the drop (increase) in the
energy of the $K^+$ ($K^0$) meson at zero momentum,
from the isospin symmetric case, is about 9 MeV (11 MeV), whereas
these become about 30 MeV (42 MeV) at a density of $\rho_B=4 \rho_0$.

For the antikaons, the $K^-(\bar {K^0})$ energy at zero momentum
is seen to increase (drop) with the isospin asymmetry parameter, $\eta$, 
The energies of the antikaons for various values of the strangeness
fraction, $f_s$ are plotted in figures 2,4 and 6.  
The sensitivity of  the isospin asymmetry
dependence of the energies is seen to be larger for $K^-$ with
density as compared to $\bar {K^0}$.
The isospin asymmetry gives rise to an increase in the mass
of the $K^-$-mesons as compared to the isospin symmetric situation
of $\eta$=0, and hence would delay the onset on $K^-$ condensation
to higher densities, inside the neutron star matter.
The energies of the antikaons as functions of densities
are plotted for $f_s$=0.1, 0.3 and 0.5 in figures 2, 4 and 6 respectively.
The changes in the masses of the $K^-$ and $\bar {K^0}$ for $f_s$=0.1,
at $\eta$=0.5 from the isospin symmetric case, are 16 MeV (14  MeV)
at $\rho_B=\rho_0$,
and at $\rho_B=4 \rho_0$ are 47 MeV (32 MeV). For strangeness
fraction, $f_s$=0.5, these values are modified to 15 MeV (12 MeV)
at $rho_B=\rho_0$ and 45 MeV (37 MeV )for $\rho_B=4 \rho_0$.

%The energies of the kaons and antikaons, with respect to the
%isospin symmetric case, for different values of the isospin
%asymmetric parameter, $\eta$ are plotted as functions of the
%momentum in figures 4 and 5.  The energies of the kaons and antikaons
%are plotted for densities $\rho_B=\rho_0$, $2\rho_0$ and $4 \rho_0$
%in the same figures. These are seen to be more sensitive
%to momentum as we increase the isospin parameter. The momentum
%dependence turns out to be stronger for higher densities, and in
%particular, the effect seems to be more significant for $K^0$ (as
%compared to $K^+$) and $K^-$ (as compared to $\bar {K^0}$).

The qualitative behavior of the isospin asymmetry dependencies of
the energies of the kaons and antikaons at finite momenta
are reflected in
their optical potentials plotted in figures 7, 9 and 11 for the kaons, 
and in figures 8, 10 and 12 for the antikaons, at selected densities, 
for strangeness fractions $f_s$ =0.1, 0.3 and 0.5 respectively.
The different
behavior of the $K^+$ and $K^0$, as well as for the $K^-$ and
$\bar {K^0}$ optical potentials in the dense asymmetric nuclear
matter should be observed in their production as well as
propagation in isospin asymmetric heavy ion collisions. In particular
an experimental study of the $K^+/K^0$ as well as $K^-/\bar{K}^0$ ratios 
as well as their collective flow in the asymmetric heavy ion collision
experiments might be promising tools to investigate the
isospin effects discussed here.
The effects of the isospin asymmetric optical potentials could thus be
observed in nuclear collisions at the CBM experiment at the
proposed project FAIR at GSI, where experiments with neutron rich
beams are planned to be implemented.

\section{Summary}

To summarize, within a chiral SU(3) model we have investigated the
density dependence of the kaon and antikaon optical potentials in
asymmetric hyperonic matter, arising from the interactions with
the baryon octet (originating from a vectorial Weinberg-Tomozawa interaction,
an isospin symmetric range term and an isospin asymmetric range term)
and due to scalar-isoscalar mesons ($\sigma$, $\zeta$)
and scalar-isovector $\delta$ mesons.
The properties of the nucleons, hyperons and scalar mesons
are modified in the hadronic medium and hence due to their interactions 
with the kaons and antikaons, modifiy the $K (\bar K)$-meson properties.
The model with
parameters fitted to reproduce the properties of hadron masses in
vacuum, nuclear matter saturation properties and low energy KN
scattering data, takes into account all terms up to the next to
leading order arising in chiral perturbative expansion for the
interactions of $K (\bar K)$-mesons with baryons. 

There is a significant density dependence of the isospin asymmetry
on the optical potentials of the kaons and antikaons.
This dependence seems to be even more dominant for larger
values of the strangeness fractions in the dense hadronic matter.
The results can be used in heavy-ion simulations that include mean fields for
the propagation of mesons \cite{kmeson}. The different potentials
of kaons and antikaons can be particularly relevant for
neutron-rich heavy-ion beams at the CBM experiment at the future
project FAIR at GSI, Germany, as well as at the experiments at the
proposed Rare Isotope Accelerator (RIA) laboratory, USA.
The $K^+/K^0$ as well as  $K^-/\bar{K}^0$ ratios as well their 
flow pattern for different isospin of projectile
and target is a promising observable to study these effects.
Furthermore, the medium modification of antikaons due to isospin
asymmetry in dense matter can have important consequences,
for example on the onset of antikaon condensation in the bulk
charge neutral matter in neutron stars. The effects of finite 
temperatures on optical potentials of kaons and antikaons
and their possible implications on high energy heavy ion collision 
experiments are under investigation.

\acknowledgements
One of the authors (AM) is grateful to the Institut
f\"ur Theoretische Physik Frankfurt for the
warm hospitality where the present work was initiated.
AM acknowledges financial support from
Alexander von Humboldt Stiftung when this work was initiated. 
The use of the resources of the
Frankfurt Center for Scientific Computing (CSC) is additionally
gratefully acknowledged.

\end{document}